%% file: 00_main_paper.tex
\documentclass[10pt,a4paper]{article}
\usepackage[T1]{fontenc}
\usepackage[utf8]{inputenc}
\usepackage{amssymb,amsmath,amsthm,amsfonts}    % math stuff
\usepackage{bm}                                 % bold math
\usepackage{mathtools}                          % tools to create operators and delimiters
\usepackage{stmaryrd}                           % additional math symbols
\usepackage{breqn}                              % automatic line breaking in equations
\usepackage[textfont=footnotesize,labelfont={footnotesize,bf}]{caption}
\usepackage{subcaption}
\usepackage{booktabs}
\usepackage[inline]{enumitem}
\usepackage{graphicx}
\usepackage[dvipsnames]{xcolor}
\usepackage[top=30mm,right=30mm,left=30mm,bottom=30mm]{geometry}
\usepackage[colorlinks=true,linkcolor=blue,urlcolor=blue]{hyperref}
\usepackage[affil-it,auth-sc]{authblk}
\usepackage[colorinlistoftodos,textwidth=25mm,textsize=footnotesize]{todonotes}
\usepackage{lipsum}                             % generate dummy text
\usepackage[section]{placeins}
\usepackage{siunitx}
\usepackage[
	backend=biber,
	style=numeric,
	citestyle=authoryear-comp,
	maxbibnames=99,
	minbibnames=99,
	maxcitenames=2,
	mincitenames=1,
	giveninits=true,
	sorting=none,
	sortlocale=auto,
	natbib=true,
	url=false,
	isbn=false,
	doi=true,
	eprint=true
]{biblatex}
\input{00_definitions}                             % useful math definitions
\graphicspath{{./}{./figures/}}                 % path to figures folder

\usepackage[linesnumbered, vlined]{algorithm2e}
\SetKwInput{KwInput}{Input}                % Set the Input
\SetKwInput{KwOutput}{Output}              % set the Output

\usepackage{xcolor}
\usepackage{tcolorbox}

\def\censura{1}

\title{A computational model for inelastic behaviour and fracture of refractory industrial components under high-temperature conditions, application to slide gate plates}

\ifnum \censura=1 {
    \author[1]{Lorenzo Fiore\footnote{Corresponding author: e-mail: \href{lorenzo.fiore@unitn.it}{lorenzo.fiore@unitn.it}; phone: +39\,0461\,282507.}}
    \author[1]{Andrea Piccolroaz}
    
    \affil[1]{Department of Civil, Environmental, and Mechanical Engineering, University of Trento, via Mesiano 77, 38123 Trento, Italy}
} \else {
    \author[1,2]{Lorenzo Fiore\footnote{Corresponding author: e-mail: \href{lorenzo.fiore@unitn.it}{lorenzo.fiore@unitn.it}; phone: +39\,0461\,282507.}}
    \author[2]{Gregory Spingler}
    \author[1]{Andrea Piccolroaz}
    
    \affil[1]{Department of Civil, Environmental, and Mechanical Engineering, University of Trento, via Mesiano 77, 38123 Trento, Italy}
    \affil[2]{Vesuvius Ghlin, Rue de Douvrain 17, 7011 Mons, Belgium}
} \fi

\date{\today}

\begin{document}

\maketitle

\begin{abstract}
\noindent
This work aims to provide a computational model that can describe the complex behaviour of refractory industrial components under working conditions. Special attention is given to the asymmetric tension-compression behaviour and its evolution in the full range of working temperatures. The model accounts for inelastic flow in compression and brittle fracture behaviour in tension by leveraging the continuum-mechanics theory of plasticity and phase-field fracture damage.
The model is implemented in the Finite Element open-source platform FEniCS and is used to analyze the fracture phenomenon in the refractory plate used in ladle slide gate systems to control the liquid steel flow from the ladle to the tundish.
\end{abstract}

\paragraph{Keywords}
Phase field fracture \textperiodcentered\
Plasticity \textperiodcentered\
Thermal shock \textperiodcentered\
Refractory Ceramics

% ------------------------------------------------------------------70
% Main content
% ------------------------------------------------------------------70

\input{01_Introduction}
\input{02_State_of_the_art}
\input{03_Model}
\input{04_Industrial_application}
\input{05_Conclusion}

% ------------------------------------------------------------------70
% Additional infos and references
% ------------------------------------------------------------------70

\section*{CRediT authorship contribution statement}
% Legend: https://beta.elsevier.com/researcher/author/policies-and-guidelines/credit-author-statement?trial=true
\textbf{Lorenzo Fiore:} Conceptualization, Methodology, Software, Writing - Original Draft.
\textbf{Andrea Piccolroaz:} Funding acquisition, Supervision, Writing - Review \& Editing.
\ifnum \censura=1 {} \else {
\textbf{Gregory Spingler:} Supervision.
} \fi

\section*{Funding}
This work was supported by funding from the European Union’s Horizon 2020 research and innovation programme under the Marie Skłodowska-Curie grant agreement No 955944 - RE-FRACTURE2 - Modelling and optimal design of refractories for high temperature industrial applications for a low carbon society. 

\section*{Declaration of Competing Interest}
The authors declare that there is no conflict of interest. 

\section*{Acknowledgements}
The model and results reported in the present document summarize the findings presented in the industrial PhD thesis of Lorenzo Fiore "Development of thermoplastic constitutive models for refractory ceramics in wide temperature range" which was realized in collaboration between the Univerisity of Trento and the company Vesuvius Ghlin.
The authors are grateful to Pietro Lenarda from IMT School for Advanced Studies Lucca Italy for the support on the use of the open-source software FEniCS and for fruitful discussions and insights about the phase-field fracture approach.
The authors are grateful to Profs. D. Bigoni and M. Paggi for fruitful discussions and useful comments.

\appendix
\input{06_Appendix}

\printbibliography

\end{document}

%% file: 00_definitions.tex
\DeclareMathOperator{\tr}{tr}

%% file: 01_Introduction.tex
\section{Introduction}

% Refractories are quasi-brittle materials 

From a continuum mechanics point of view, ceramic materials can be classified as quasi-brittle granular materials, a class of materials that also includes soil, rocks, concrete, porous metals and general granular media \citep{Anderson_2005}. The terminology clearly highlights the main characteristic of the structure of these materials: the presence of heterogeneous grains suspended in a homogeneous matrix. The usual approach in modelling and numerical simulation is that of describing the composite nature of the material from a phenomenological point of view: macroscopically, the material behaviour is characterized by an elastic reversible response up to a certain stress threshold followed by irreversible effects of different nature depending on the kind of load, compressive or tensile, so that the material can be described using a homogenized approach that neglects the specific microstructure.

% Refractory behavior in compression and relative models

Experimental observations \citep{Blond_Schmitt_Hild_Blumenfeld_Poirier_2005, Bareiro_deAndradeSilva_Sotelino_Gomes_2018} on the response of ceramic refractories under compression show that, upon reaching a specific limit load, the materials undergo inelastic flow characterized by a progressive monotonic decrease of the reaction response and progressive accumulation of permanent deformation. In the context of continuum mechanics, this kind of behaviour can be described using the theory of incremental plasticity \citep{Simo_Hughes_1998}. Since the theory of plasticity was initially developed to represent the behaviour of metals from a rigorous point of view, its application to granular materials is unjustified \citep{Lubliner_Oliver_Oller_Onate_1989} such an approach is thus only phenomenological, without any pretence to explain the phenomena happening at the micro-scale. Nonetheless, its application has yielded satisfactory results: in \cite{Andreev_Harmuth_2003}, a Drucker-Prager yield surface is used to determine the onset of plastic flow. At the same time, a Rankine stress cap signals the limit tensile load, the resulting model is used to study the thermo-mechanical behaviour of the refractory lining of a teeming ladle. The combination of two different limit criteria is further developed in \cite{Brannon_Fossum_2004, Fossum_Brannon_2006, Foster_Regueiro_Fossum_Borja_2005} with the formulation of the yield surface as a closed surface. In these models, the yield surface can change in shape and size to represent the evolution of the material properties with hardening. In contrast, the failure limit is modelled with an additional surface. 

% Limit of applicability of plasticity models. Refractories in tension

The application of the theory of plasticity to describe ceramic refractories is hindered by the fact that under tensile load, these materials show loss of cohesion and fracture both at the micro-scale and macroscopically (experimental observations can be found in \citep{Schmitt_Berthaud_Poirier_2000, Nazaret_Marzagui_Cutard_2006, Teixeira_Samadi_Gillibert_Jin_Sayet_Gruber_Blond_2020}). In plasticity models, reaching the limit surface does not imply failure at the material point but rather signals the limit of applicability of the constitutive models for describing the stress state. This argument is reported in \cite{Brannon_Fossum_2004} specifically for a material model designed for granular materials. Still, it is valid for all local material models \citep{de_Borst_Sluys_Muhlhaus_Pamin_1993}: no local material model can represent localization and structural failure as those phenomena involve loss of validity of the regular equilibrium PDE. 

% How to actually model fractures

The general hypothesis in continuum mechanics is that of continuity of the unknown displacement field, this hypothesis does not hold in the explicit representation of fracture, as fracture represent an explicit discontinuity. Accurate modeling of fracture phenomena thus requires ad-hoc expedients, a comparative study of the main models available to describe fracture in quasi-brittle materials can be found in \cite{Cervera_Barbat_Chiumenti_Wu_2022}, this work is focused on gradient-damage models.

% From Peerling to Damhof

In non-local models, loss of stiffness at a specific material point is computed considering the stress state at the surrounding points. In \cite{Peerlings_De_Borst_Brekelmans_De_Vree_1996} this approach is implemented by computing a non-local variable as integral over the domain multiplied by a weighting function, this formulation results in an additional PDE for the computation of the non-local variable and it is shown to yield consistent results independently from the discretization mesh. The application of non-local damage models for the study of quasi-brittle granular materials can be traced back to the work of \citeauthor{Comi_1999} \citep{Comi_2001, Comi_1999}, which is used as a reference for the application study of \cite{Schmitt_Burr_Berthaud_Poirier_2002} on refractory ceramics components. In \cite{Özdemir_Brekelmans_Geers_2010, Damhof_Brekelmans_Geers_2011} the gradient-damage formulation from \cite{Peerlings_De_Borst_Brekelmans_De_Vree_1996} is enriched with additional terms related to the evolution of the mechanical response of the material with the evolution of temperature, the resulting model is applied to study refractory installations. 

% Damhof and phase-field-fracture

As with \cite{Peerlings_De_Borst_Brekelmans_De_Vree_1996}, the model of \cite{Damhof_Brekelmans_Geers_2011} presents a dedicated PDE for the modelling of localization of damage. Interestingly, the aforementioned PDE has the same mathematical structure as the one appearing in phase-field fracture models, which have enjoyed considerable success in recent years as models for fracture phenomena. With the phase-field approach \citep{Francfort_Marigo_1998, Bourdin_Francfort_Marigo_2000}, the fracture discontinuity is represented as a level set of a continuous function. This continuous function can be thought of as damage, which is thus not a local property but a field with its own PDE derived in a variational form based on energetic considerations based on the fundamental observations of \cite{Griffith_1920}. The similarities between the approach of \cite{Damhof_Brekelmans_Geers_2011} and the standard phase-field fracture formulation encourage the application of this theory to the fields of numerical simulation of refractories, since the authors specialized in this field arrived at the same mathematical formulation with a different approach from the pure variational one of \cite{Francfort_Marigo_1998}. 

% Introduction: the gap

Numerous contributions in the literature present constitutive models designed to capture crack initiation and propagation in porous media and mixed thermal-mechanical loads, and among these works \cite{Choo_Sun_2018} and \cite{Lenarda_Reinoso_Paggi_2022} deserve particular attention for the comprehensive approach which takes traction and compression separately into account as crack drivers in phase-field fracture formulations. In all these cases, though, the presented results feature test geometries rather than real components. 

% Introduction: filling the gap

This contribution presents a thermo-plastic material model for ceramic refractory materials based on the work of \cite{Poltronieri_Piccolroaz_Bigoni_Romero_Baivier_2014} on the constitutive modelling of granular materials and enriches the formulation with a stress-based phase-field fracture model following the developments of \cite{Miehe_Schänzel_Ulmer_2015}. The model is equipped with evolution laws for the material properties that take macroscopically into consideration the variation of material properties due to thermo-chemical changes in the material microstructure. For validation and testing purposes, the model has been implemented in the open-source finite-element framework FEniCS, and it has been used to study the fracture phenomenon occurring in operational conditions on Slide Gate Plates, an industrial component of relevant importance in steel plants for controlling the flow of molten metal. The improvement of the design of the Slide Gate Plate is an active area of research in the industry. In \cite{Lee_Thomas_Kim_2016}, the root causes of plate fracture are analyzed using a simple elastic material model and observing the principal stress state. In contrast, \cite{Tang_Lu_Li_Zhu_Yi_Liu_Eckert_2023} uses the Concrete Damage Plasticity model available in the Abaqus software \citep{ABAQUS} to infer fracture path and propose optimizations of the design, based on distributions of different materials. Using the approach proposed in the present contribution, it is possible to study actual fracture initiation and propagation in the piece.

% Structure of the paper

The following document is divided into three sections: the first part is dedicated to the theory of gradient-damage mechanics with a presentation of the phase-field approach to fracture and an analysis of the similarities with the approach of \cite{Damhof_Brekelmans_Geers_2011}; in the second part the novel constitutive model for refractory ceramic is presented and the design choices taken are justified along with some details of implementation; the third part is dedicated to the industrial application with a description of the application case and the thermo-mechanical model used to represent the working condition of Slide Gate Plates in steel plant.

%% file: 02_State_of_the_art.tex
\section{Gradient damage models for refractory ceramics}

The standard phase field formulation of fracture is rooted in the original observation of \cite{Griffith_1920}: crack propagation can happen only when the available internal energy overcomes the energy required to create a new crack surface. The expended energy per unit of newly opened crack surface is a measurable material property, with an experiment such as the Wedge Splitting Test \citep{Brühwiler_Wittmann_1990, Harmuth_1995, Stueckelschweiger_Gruber_Jin_Harmuth_2019}, and it is usually referred to as $G_c$. Leveraging this idea, it is possible to account for the possibility of the insurgence of a crack in a continuum from the variational formulation of the structural problem. Denoting with $\Pi$ the total potential energy of a mechanical system associated with the continuum body $\Omega$:
\begin{equation}
    \Pi = \int_\Omega \psi\, d\Omega + \int_\Gamma G_c\, d\Gamma - \int_\Omega \bm{b} \cdot \bm{u}\, d\Omega - \int_{\partial \Omega} \bm{t} \cdot \bm{u}\, d\partial\Omega,
\end{equation}
where $\psi$ is the internal energy per unit volume, $\Gamma$ is the crack surface, while $\bm{b}$ and $\bm{t}$ are respectively volume and surface traction external forces energetically conjugate to the system displacement vector field $\bm{u}$. Performing the above integration requires knowledge of the evolution of the crack surface $\Gamma$ as a geometrical domain, which is one of the main difficulties in the numerical treatment of fractures. Phase field methods overcome this problem by introducing a damage field variable $d$ and a related crack surface density $\gamma(d, \nabla d)$. Using these concepts, the second energy integral can be rewritten as:
\begin{equation}
    \int_\Gamma G_c\, d\Gamma = \int_\Omega G_c\, \gamma(d, \nabla d)\, d\Omega.
\end{equation}
In this way, integration over the unknown domain $\Gamma$ is substituted with an integration over the fixed domain $\Omega$. The choice of $\gamma(d, \nabla d)$ is a constitutive choice, and different formulations are available. The one mostly used in the literature is that of \cite{Ambrosio_Tortorelli_1990}:
\begin{equation}
    \gamma(d, \nabla d) = \dfrac{1}{2} \left( \dfrac{d^2}{\ell} + \ell |\nabla d|^2 \right),
\end{equation}
where $\ell$ is a length scale parameter used to regularise the sharp crack domain. The above expression for $\gamma(d, \nabla d)$ respects the convergence to the sharp interface $\Gamma$ in the limit for $\ell \rightarrow 0$. Using this approximation, the total potential energy $\Pi$ can be rewritten as:
\begin{equation}\label{integral_Pi}
    \Pi = \int_\Omega \psi\, d\Omega + \int_\Omega  \dfrac{G_c}{2} \left( \dfrac{d^2}{\ell} + \ell |\nabla d|^2 \right) d\Omega - \int_\Omega \bm{b} \cdot \bm{u}\, d\Omega - \int_{\partial \Omega} \bm{t} \cdot \bm{u}\, d\partial\Omega.
\end{equation}
The above expression clearly shows that in this formulation, an energetic cost is associated with a non-zero value of the damage variable in the domain $\Omega$ and a non-zero value of the gradient of $d$. Both contributions are weighted using the length scale parameter $\ell$ so that as $\ell$ increases, the overall structure response is stiffer because a greater energetic cost is associated with damage development. In contrast, as $\ell$ decreases, damage localization is promoted.

To model the decrease in stiffness caused by the insurgence of damage, a degrading function $g(d)$ can be multiplied by the elastic strain energy. The function $g(d)$ has to assume a value of 1 in the case of undamaged material and 0 in the case of fully damaged material. The simple expression $g(d) = (1-d)^2$ meets these requirements, and it also presents a non-zero slope at $d=0$, preventing possible instabilities in the case of abrupt stiffness drop.

\cite{Miehe_Welschinger_Hofacker_2010} also propose to account for traction-compression asymmetry and crack closure effects by splitting the strain energy into a degraded and a residual stored part:
\begin{equation}
    \psi = \psi_D + \psi_R.
\end{equation}
This additional feature of the model helps represent the physical observation that crack initiation is generally associated with tensile stresses only and, after cracking, the material cannot withstand tensile stresses and thus cannot accumulate tensile strain energy. Still, it can withstand compressive stresses and accumulate the related energy. In this sense, $\psi_D$ and $\psi_R$ represent tensile and compressive strain energies, respectively. A possible formulation to represent this is the spectral decomposition proposed by \cite{Miehe_Welschinger_Hofacker_2010}:
\begin{equation}
    \begin{aligned}
        \psi_R(\bm\varepsilon_e) &= \dfrac{\lambda}{2} [ \langle \tr(\bm\varepsilon_e) \rangle^+ ]^2 + \mu\, \tr[(\bm\varepsilon_e^+)^2], \\[2ex]
        \psi_D(\bm\varepsilon_e) &= \dfrac{\lambda}{2} [ \langle \tr(\bm\varepsilon_e) \rangle^- ]^2 + \mu\, \tr[(\bm\varepsilon_e^-)^2], \\[2ex]
        \bm\varepsilon^{\pm} &= \sum_i \langle \varepsilon^i \rangle^\pm ~ \bm{n}^i \otimes \bm{n}^i,
    \end{aligned}
\end{equation}
where $\langle \cdot \rangle^\pm$ denotes the positive and negative parts
\begin{equation}
    \langle x \rangle^\pm = \dfrac{1}{2} (|x| \pm x).
\end{equation}

This constitutive choice provides an effective representation of the physical behaviour for most materials prone to cracks, but it introduces further non-linearities in the problem and indeed, other constitutive choices are available (the volumetric-deviatoric split by \cite{Amor_Marigo_Maurini_2009} and the no-tension split by \cite{Freddi_Royer-Carfagni_2010} are two other possible choices). Recently \citep{Vicentini_Zolesi_Carrara_Maurini_Lorenzis_2024} reviewed the currently available choices for strain energy splits addressing the physical representativeness. Using the strain energy split and introducing the degrading function $g(d)$, the integral $\Pi$ \eqref{integral_Pi} can be rewritten as:
\begin{equation}
    \begin{gathered}
    \Pi = \int_\Omega \left[ g(d)\,\psi_D + \psi_R \right]\, d\Omega + \int_\Omega \dfrac{G_c}{2} \left( \dfrac{d^2}{\ell} + \ell |\nabla d|^2 \right) d\Omega + \\ - \int_\Omega \bm{b} \cdot \bm{u}\, d\Omega  - \int_{\partial \Omega} \bm{t} \cdot \bm{u}\, d\partial\Omega.
    \end{gathered}
\end{equation}
The equilibrium of the mechanical system associated with the energy integral $\Pi$ is found in the condition that the variational derivative of $\Pi$ is null. Choosing the system displacement vector field $\bm{u}$ as the unknown, the variational derivative $D_v \Pi(\bm{u})$ can be written as:
\begin{equation}
    D_v \Pi(\bm{u}) = \left[ \dfrac{d}{dh} \Pi(\bm{u} + h\, \bm{v}) \right]_{h=0},
\end{equation}
where the function $\bm{v}$ is homogeneous to the trial function $\bm{u}$ and acts as a test function. Denoting with $\delta\bm{u}$ a small variation of the function $\bm{u}$, it is possible to adopt another notation and treat $\delta\bm{u}$ as the test function increment. In the case at hand, the displacement field $\bm{u}$ is not the only unknown as the scalar damage field $d$ has to be evaluated as well, so the stationary condition reads:
\begin{equation}
    \dfrac{\partial \Pi}{\partial \bm{u}} \delta\bm{u} + \dfrac{\partial \Pi}{\partial d} \delta d = 0.
\end{equation}
The solution of the structural problem in the form reported above presents consistent challenges from a numerical point of view: the problem is non-linear, and, in the eventual choice of an implicit solution scheme, it has to be solved with incremental-iterative methods such as the Newton-Raphson method. The related stiffness matrix is not symmetrical, implying a high cost in computation and memory requirements. To overcome these difficulties, it is possible to assume, at this stage, that the problem has to be solved incrementally so that in the increment, the two unknown fields, $\bm{u}$ and $d$, can be considered independent. Under this hypothesis, it is possible to split the displacement-damage problem into two coupled problems:
\begin{align}
    \label{eq:principle of virtual work}
    \delta d &= 0 \quad \rightarrow \quad \int_\Omega \bm\sigma(d) \cdot \bm\varepsilon(\delta\bm{u})\, d\Omega - \int_\Omega \bm{b} \cdot \delta\bm{u}\, d\Omega - \int_{\partial \Omega} \bm{t} \cdot \delta\bm{u}\, d\partial\Omega = 0, \\
    \delta \bm{u} &= 0 \quad \rightarrow \quad \int_\Omega \Big\lbrace \dfrac{\partial }{\partial d} ~ \left[ g(d) \psi_D \right] + G_c \dfrac{\partial \gamma}{\partial d} \Big\rbrace \delta d\, d\Omega = 0.
\end{align}
The displacement problem is thus reconducted to the standard virtual work principle form, while the damage problem can be rewritten as:
\begin{equation}
    \int_\Omega \left[ g'(d) \dfrac{\psi_D}{G_c} \delta d + \frac{d}{\ell} \delta d + \ell\, \nabla d \cdot \nabla \delta d \right] d\Omega = 0.
\end{equation}
The above model can represent the insurgence of a crack from energetic considerations. Still, it does not take into account the irreversibility of the phenomenon. Without further modifications, the model predicts the disappearance of a crack at the decrease of the energy necessary to open it. One possible artifice to overcome this limitation is that proposed by \cite{Miehe_Welschinger_Hofacker_2010} and consists of isolating the crack driving force term and artificially imposing its monotonicity by substituting it with its maximum in time: 
\begin{gather}
    \int_\Omega \left[ g'(d)\, \mathcal{H}\, \delta d + \frac{d}{\ell} \delta d + \ell\, \nabla d \cdot \nabla \delta d \right] d\Omega = 0, \label{standard_phase_field_fracture} \\[2ex] 
    \mathcal{H} = \max_t \Big \lbrace \dfrac{\psi_D}{G_c} \Big \rbrace.
\end{gather}
This reformulation of the problem also separates two different contributions in the damage stationarity condition: the driving term, which takes as input the strain energy from the displacement problem, and the geometric resistance term involving the damage variable, its gradient and the length scale parameter solely.

% Phase-field can be thought of as a gradient-damagage model, use this as a link to Damhof
Observing expression \eqref{standard_phase_field_fracture}, it can be said that the phase-field fracture formulation can be thought of as a non-local gradient damage model with a dedicated PDE for the computation of damage. This kind of mathematical structure also appears in other gradient-damage models whose derivation differs from the phase-field fracture, such as the model from \cite{Damhof_Brekelmans_Geers_2008}. 

The non-local damage approach is applied in \cite{Damhof_Brekelmans_Geers_2011} to study the refractory lining of a teeming ladle, and it is shown to yield satisfactory results. The model is a thermo-elastic damage model in small strain; the strain is supposed to be additively decomposable into elastic and thermal, and the elastic behaviour is affected by isotropic damage $d$ by a first-order degradation function:
\begin{equation}
    \bm\sigma = (1-d)~\mathbf{D}(\bm\varepsilon - \bm\varepsilon_t) .
\end{equation}
The damage is assumed to be dependent on both mechanical and thermal effects through the linear expression:
\begin{equation}
    d = 1 - (1-d_{e})(1-d_{th}),
\end{equation}
where $d_{e}$ and $d_{th}$ are the mechanical and thermal contributions to the damage, respectively. The evolution laws for these parameters are formulated in incremental form as:
\begin{gather}
    \dot{d}_{e} = \begin{cases}
        A \langle \dot{\Bar{\varepsilon}}_{eq} \rangle ~  \exp ( - \beta d_{e} ) ~ ~ ~ \text{if} ~ ~ ~ \Bar{\varepsilon}_{eq} > \kappa_0 ~ ~ ~ \text{and} ~ ~ ~ d_{e} < 1 \\
        0 ~ ~ ~\text{otherwise}
    \end{cases} ,\\
    \dot{d}_{th} = \begin{cases}
        B \langle \dot{T} \rangle ~  \exp ( - \gamma d_{th} ) ( 1 - d_{th}^\phi ) ~ ~ ~ \text{if} ~ ~ ~ T > \kappa_0 ~ ~ ~ \text{and} ~ ~ ~ d_{th} < 1 \\
        0 ~ ~ ~\text{otherwise}
    \end{cases} ,
\end{gather}
where $A$, $B$, $\beta$, $\gamma$, $\phi$ and $\kappa_0$ are material parameters while $\Bar{\varepsilon}_{eq}$ is a non-local equivalent strain.

In \cite{Peerlings_De_Borst_Brekelmans_De_Vree_1996}, the non-local quantity $\Bar{\varepsilon}_{eq}(\bm{x})$ is calculated as an integration of a local quantity $\varepsilon_{eq}(\bm{x})$ multiplied by a smoothing function and integrated over the domain:
\begin{equation}
    \Bar{\varepsilon}_{eq}(\bm{x}) = \int_\Omega f(\xi) ~ \varepsilon_{eq}(\bm{x} + \xi) ~ d\Omega .
\end{equation}
The actual integration is avoided by approximation through Taylor expansion; this results in a formulation involving only the Laplacian:
\begin{equation}\label{eq:strong form of local quantity}
    \Bar{\varepsilon}_{eq} - \ell^2 ~ \nabla^2 \Bar{\varepsilon}_{eq} = \varepsilon_{eq}  .
\end{equation}
In this last expression, the quantity $\ell$ is the result of the integration of the product of the weight function and the Taylor expansion terms of the local quantity. The quantity $\ell$ has the dimensions of length and can be interpreted as an internal length scale.

The implicit definition in equation \eqref{eq:strong form of local quantity} leads to the definition of a PDE for the calculation of the non-local quantity $\Bar{\varepsilon}_{eq}$. The mentioned equation in weak integral form reads:
\begin{equation}
    \label{eq:Damnof non local}
    \int_\Omega (\Bar{\varepsilon}_{eq} ~ \delta \Bar{\varepsilon}_{eq} + \ell^2 ~ \nabla \delta \Bar{\varepsilon}_{eq} \cdot \nabla \Bar{\varepsilon}_{eq} ) d\Omega = \int_\Omega c(\bm{u}, T) \delta q d\Omega,
\end{equation}
where $c(\bm{u}, T)$ is a source term that in \cite{Damhof_Brekelmans_Geers_2011} is supposed to depend both on elastic strain and temperature on the hypothesis that different damage mechanisms act at different scales and specifically at the micro-scale damage is due to mismatch in thermal expansion coefficients between grains and matrix:
\begin{equation}
    c(\bm{u}, T) = \varepsilon_{eq}(\bm{u}) + \frac{C_{ths}~\rho~c_p}{K} |\dot{T}|,
\end{equation}
where $C_{ths}$ is an additional material parameter, $\rho$ is the material density, $c_p$ is the specific heat, $K$ the thermal conductivity.

As regards the expression for the local quantity $\varepsilon_{eq}$, in the \cite{Damhof_Brekelmans_Geers_2011} model the authors use the expression proposed by \cite{Peerlings_1999}:
\begin{equation}\label{eq:local quantity}
    \varepsilon_{eq} = \frac{\eta - 1}{ 2\eta (1-2\nu) }J_1 + \frac{1}{2\eta} \sqrt{\left( \frac{\eta - 1}{1 - 2 \nu} \right)^2~J_1^2 + \frac{12\eta}{(1+\nu)^2}J_2} ,
\end{equation}
where $\eta$ is the ratio of the compressive and tensile material strength, $\nu$ is the Poisson ratio while $J_1$ and $J_2$ are invariants of the elastic tensor:
\begin{gather}
    J_1 = \tr(\bm\varepsilon_{e}) ,\\
    J_2 = \tr(\bm\varepsilon_{e} \cdot \bm\varepsilon_{e}) - \frac{1}{3} \tr^2(\bm\varepsilon_{e}) . 
\end{gather}

Interestingly enough, the formulation reported in equation \eqref{eq:Damnof non local} is equivalent in its structure to the phase-field fracture formulation as the non-local quantity $\Bar{\varepsilon}_{eq}$ can be replaced with damage $d$ and the source term $c(\bm{u}, T)$ with the crack driving force $\mathcal{H}$. The model from \cite{Damhof_Brekelmans_Geers_2008} implements the general principles of non-locality, monotonicity and irreversibility of damage and also, the concept of a damage driving force is implemented in some form through the expression for the local equivalent quantity. In the evolution expressions for the damage, the model requires several parameters that do not have a direct physical meaning. The non-local quantity equation \eqref{eq:Damnof non local} instead only requires the parameter $\ell$, which has a direct interpretation as a material characteristic length, whose validity range lies between the minimum length necessary to observe homogenization of the material behaviour and the maximum size of localized effects such as fractures.

The model contains all the ingredients of a phase field fracture model even if the derivation does not start from simple variational principles as in \cite{Francfort_Marigo_1998}.

%% file: 03_Model.tex
\section{Constitutive model for ceramic refractory materials}

Ceramic materials are quasi-brittle materials in iso-thermal conditions at the reference temperature. However, in the specific loading conditions to which refractory products are subjected in operation, significant changes in the thermo-chemical properties happen in the material so that, depending on the specific composition, the material behaviour can deviate from brittleness.
In this context, failure is triggered either by mechanical stresses induced by non-homogeneous thermal loading of the piece, by complete loss of strength due to the high temperature, or by a combination of the two.

Capturing all those effects in a material model poses some challenges. Such a model has to be designed to account for:
\begin{itemize}
    \item inelastic irreversible deformation;
    \item hardening and softening of the material properties induced by stress and temperature;
    \item the effect of hydrostatic pressure on the mechanical failure of a material point;
    \item the effect of temperature on the failure of a material point.
\end{itemize}

A suitable choice as a starting point for the development of a constitutive model that can suit these requirements is offered by the incremental theory of elasto-plasticity. Numerous models are already available in the literature to describe the behaviour of quasi-brittle granular materials, in particular concrete and rock, and among those, the works of \cite{Foster_Regueiro_Fossum_Borja_2005} and \cite{Fossum_Brannon_2006} focus on the stress triaxiality effect on failure, proposing failure criteria specifically designed for pressure sensitive materials. 

Due to the design choices and the specific approach used to implement these choices, the cited models lack the flexibility to adapt to different variations of the same material. In contrast, this flexibility is quite important in the refractory industry, considering the continuous improvement of material recipes for function purposes and technological improvements. The model proposed by \cite{Poltronieri_Piccolroaz_Bigoni_Romero_Baivier_2014} overcomes this limitation with a constitutive model based on the highly flexible Bigoni-Piccolroaz yield criterion \citep{Bigoni_Piccolroaz_2004} and a class of simple isotropic hardening laws for the mechanical hardening. The following presents the detailed formulation for the proposed thermo-plastic refractory material model. 

The model proposed in this contribution is formulated in small strains, which is consistent with what is observed experimentally and in industry. Denoting by $\bm{u}$ the displacement field, deformation is computed using the measure:
\begin{equation}
	\bm\varepsilon(\bm{u}) = \frac{1}{2} \left( \nabla \bm{u} + \nabla \bm{u}^T \right).
\end{equation}

Assuming a linear combination of effects, deformation is additively split into elastic, plastic and thermal contributions, whose evolution is tracked separately:
\begin{equation}
	\bm{\varepsilon} = \bm{\varepsilon}_e + \bm{\varepsilon}_p + \bm{\varepsilon}_t.
\end{equation}

\subsection{Elastic reversible deformation}
Following standard thermodynamics derivations \citep{Coleman_Noll_1963, Coleman_Gurtin_1967, Bigoni_2012}, the stress state of a material point is calculated as the tensor derivative of a scalar quantity, the elastic strain energy $\psi$, with respect to the elastic strain $\bm\varepsilon_e$:
\begin{equation}
	\bm\sigma = \frac{\partial\psi}{\partial\bm\varepsilon_e}.
\end{equation}

The choice of the expression of $\psi$ fundamentally impacts the constitutive model as it enters directly into the energy balance equation to solve a thermo-mechanical loading problem. Different choices are available depending on the need to model linear or non-linear elastic behaviour and thermal effects. The model proposed in this contribution features a quadratic elastic potential formulated directly on elastic strains where temperature affects the elastic parameters:
\begin{equation}
    \psi(T, \bm\varepsilon) = \dfrac{\lambda(T)}{2} \tr(\bm\varepsilon_e^2) + 
    \mu(T)\, \bm\varepsilon_e^2,
\end{equation}
where $\lambda(T)$ and $\mu(T)$ are the Lamé parameters assumed to be thermal dependent:
\begin{gather}
    \lambda(T) = \dfrac{E(T)\, \nu}{(1+\nu)(1-2\nu)}, \\
    \mu(T) = \dfrac{E(T)}{2(1+\nu)} .
\end{gather}
The thermal dependency of the elastic parameters is assumed to be a function only of the thermal dependence of the Young modulus, and it is tracked by piece-wise linear interpolation of experimental data, because this parameter is relatively easy to measure, even in high-temperature case scenarios, and it has a wide variation influenced by chemical reactions \citep{Warchal_Andre_DeBastiani_Guillo_Huger_Martelli_Mazerat_Romero-Baivier_2017}.

\subsection{Yield criterion}
The evaluation of the severity of a stress state in a material point is performed using a scalar function, the yield criterion, that takes as input the information about the stress, either in tensor form, in terms of stress invariants or hydrostatic-deviatoric components. The material is considered to be in an elastic regime if the scalar value is negative. Irreversible deformations and failure occur otherwise. The definition of such a function must consider the effects of stress triaxiality and the dependence of failure on the hydrostatic pressure. 

The Bigoni-Piccolroaz yield function (\cite{Bigoni_Piccolroaz_2004}, referred to from here on as BP yield function)  is a flexible yield criterion represented by a closed drop-like shape in the principal stress space and formulated in terms of hydrostatic pressure $p$, deviatoric stress $q$ and Lode's angle $\theta$:
\begin{equation}
	F(\bm\sigma) = f(p) + \dfrac{q}{g(\theta)}, 
\end{equation}
where
\begin{gather}
    \begin{aligned}
        f(p) &= 
        \begin{cases}
		      - M p_c \sqrt{(\Phi-\Phi^m)[2(1-\alpha)\Phi+\alpha]} & \Phi \in [0,1] \\[2mm]
		      + \infty & \Phi \notin [0,1]
	       \end{cases}, \\
	g(\theta) &= \left( \cos\left[ \beta \dfrac{\pi}{6} - \dfrac{\cos^{-1} (\gamma \cos 3\theta)}{3} \right]  \right)^{-1}, 
    \end{aligned}
\end{gather}
are the `meridian' and `deviatoric' functions, respectively,
\begin{equation}
    \label{eq: Phi}
    \Phi = \dfrac{p + c}{p_c + c} ,
\end{equation}
and the stress invariants are defined as
\begin{equation}
    \begin{gathered}
        p = - \dfrac{\tr(\bm\sigma)}{3}, \quad
        q = \sqrt{3 J_2}, \quad
        \theta = \dfrac{1}{3} \arccos\left( \dfrac{3\sqrt{3}}{2} \dfrac{J_3}{J_2^{3/2}} \right), \\
        J_2 = \dfrac{1}{2} \tr(\bm{s}^2), \quad
        J_3 = \dfrac{1}{3} \tr(\bm{s}^3), \quad 
        \bm{s} = \bm\sigma - p \bm{I}.
    \end{gathered}
\end{equation}
As explicitly indicated by the formulation, in yield the function $F(\bm\sigma)$, the contributions of the stress representative parameters $p$, $q$ and $\theta$ are separated in different subfunctions. This yield criterion features seven material parameters which appear in separate subfunctions, so that it is possible to isolate the effect of each parameter on the predicted yield behaviour. Parameters $M$, $p_c$, $c$, $m$, and $\alpha$ dictate the shape of the yield surface in a $p-q$ plane (a plane containing the hydrostatic axis, the tri-sector of the first octant of the principal stress space), parameters $\beta$ and $\gamma$ dictate the shape in a deviatoric section. Among these parameters, $p_c$ plays a fundamental role, because it is directly related to the material resistance in a case of hydrostatic compression.

The variation of the seven parameters of the BP yield criterion during loading, through appropriate hardening laws, can represent the isotropic hardening of the material. The formulation is not directly suited to represent kinematic hardening, even though that would imply only minor modifications, and the effect is not of interest in the case of ceramic refractories.

\subsection{Incremental irreversible deformation}
The basic idea behind the continuum mechanics plasticity theory is that the stress state of a continuum body at any given time depends not only on the current state of deformation at that time but also on the whole deformation history. The use of field variables tracks this history. 

The history field variables evolution equations are stated in incremental terms:
\begin{equation}
	\dot{\bm\varepsilon}_p = \dot{\gamma}\, \bm{N}(\bm\sigma, \bm{A}),
\end{equation}
where the scalar $\gamma$ is named plastic multiplier, its increment $\dot{\gamma}$ gives the modulus of the plastic strain increment, while $\bm{N}(\bm\sigma, \bm{A})$ is a tensor carrying the information for the plastic strain direction. The tensor $\bm{N}(\bm\sigma, \bm{A})$ depends not only on the stress state $\bm\sigma$ but also on the vector $\bm{A}$ containing the updated values of the hardening variables, in this case, the seven parameters of the BP yield criterion.

The critical point at this stage is in the choice of the evolution law for the $\bm{N}(\bm\sigma, \bm{A})$ tensor. The usual choice is linking the plastic strain evolution with the failure criterion using the hypothesis that the thermodynamically admissible plastic deformation results in the largest possible energy dissipation. This criterion is respected if the scalar product between the plastic strain increment tensor and the stress increment tensor is maximum. Leveraging the representation through the principal stress space and the yield surface, the maximum plastic dissipation criterion can be reformulated with the geometrical condition that the plastic strain direction tensor has to be the derivative of the yield function with the stress:
\begin{equation}
    \label{eq:associative flow rule}
	\dot{\bm\varepsilon}_p = \dot{\gamma} \frac{\partial F(\bm\sigma)}{\partial \bm{\sigma}}.
\end{equation}
This hypothesis is referred to as the associative hypothesis, and it is the one followed in this contribution. 

In order to track the severity of the irreversible deformations, the material point is subjected to during loading, the common practice is to introduce a scalar quantity, the accumulated plastic strain $\bar{\varepsilon}_p$:
\begin{equation}
    \bar{\varepsilon}_p = \int_0^t ||\dot{\bm\varepsilon}_p||\, dt.
\end{equation}
The corresponding increment of accumulated plastic strain is calculated as:
\begin{equation}
    \dot{\bar{\varepsilon}}_p = \dot{\gamma}\, ||\bm{N}(\bm\sigma, \bm{A})||.
\end{equation}

\subsection{Mechanical hardening laws}
In order to represent the effect of material hardening due to mechanical and thermal laws, the parameters defining the yield criterion are subjected to evolution equations in which the accumulated plastic strain variable and the temperature enter as inputs.

Using the BP yield criterion, it is possible to modify the volume and shape of the yield surface separately by modifying the values of specific parameters. In particular, in the most straightforward representation of isotropic hardening, the shape of the yield surface is fixed in the $p-q$ and deviatoric planes. Hence, the only parameters subjected to variation due to loading are $c$ and $p_c$. In practice, experimental observations show a consistent variability in the value of the $p_c$ parameter, so that it is acceptable to neglect the variation of the $c$ parameter.

With this last hypothesis, the isotropic hardening of the ceramic material is represented by the sole variation of parameter $p_c$. An additional hypothesis is related to the combination of the mechanical and thermal effects, which is supposed to be linear so that:
\begin{equation}
    p_c = p_{cT} + p_{cM}.
\end{equation}
The mechanical effect of isotropic hardening in granular materials is well captured by the class of laws proposed by \cite{Poltronieri_Piccolroaz_Bigoni_Romero_Baivier_2014}. In this contribution, the choice of a specific law from the available class falls on the linear case:
\begin{equation}
    p_{cM} = \dfrac{\chi}{1 + \delta\, \bar\varepsilon_p} \bar\varepsilon_p, 
\end{equation}
where $\chi$ and $\delta$ are additional material parameters.

In the loading scenario, capturing the evolution of the material properties with temperature is essential. Indeed, failure at high temperatures for loss of material integrity is a possible risk. 

The main problem related to modelling mechanical properties in the case of such high-temperature variations is that material chemistry plays an important role. It is well-known experimental evidence the fact that material properties change consistently at the onset of a certain specific temperature that depends on the specific material \citep{Warchal_Andre_DeBastiani_Guillo_Huger_Martelli_Mazerat_Romero-Baivier_2017}. Given the complexity of the phenomena involved, it is not feasible to use an analytic expression to represent all kinds of ceramic refractories this model targets. For this reason, the strategy adopted here allows the dependence of $p_{cT}$ on temperature using an interpolation of experimental data.

\subsection{Visco-plasticity}
The general theory of plasticity is formulated for quasi-static solutions in reasonably low strain rate conditions. In this context, the introduction of incremental equations does not necessarily imply the introduction of physical time. However, in the case of the thermo-mechanical loading of refractory components, time scale can play a role. Depending on the specific application, the instant a refractory component has its first contact with the liquid steel can be sudden, and the consequent thermal loading can be abrupt. In these conditions, the elastic response of a material is immediate, while irreversible effects take a finite amount of time to manifest. This statement can be justified by the consideration that internal frictional sliding and pore collapse happen at a different time scale compared to elastic wave propagation in a solid. 

This line of reasoning brings to the hypothesis that, in conditions of high strain rates, a material can withstand a higher amount of stress compared to the quasi-static loading conditions, but only for a brief amount of time, after which the stress state relaxes, and the material comes back to the quasi-static solution. The over-stress time is considered an additional material property, possibly influenced by other factors such as temperature. 

A simple way to implement these considerations in an elastoplastic material model consists of relaxing the constraint that the stress state has to lie on the yield surface and allowing stress states out of the yield surface. This implies the addition of a new relationship for the calculation of the stress state. In \citeauthor{Perzyna_1963} formulation \citep{Perzyna_1963, Perzyna_1966, Perzyna_1971} such a relationship is stated in terms of an explicit formula for the plastic multiplier $\gamma$:
\begin{equation}\label{explicit definition plastic multiplier}
    \dot{\gamma} = \eta \langle F(\bm\sigma, \bm{A}) \rangle .
\end{equation}
In \citeauthor{Perzyna_1963} formulation, $\eta$ is an additional material parameter, while the formulation proposed by \cite{Duvaut_Lions_1972} is better suited to represent the time-related effects:
\begin{equation}
    \eta(\Delta t) = \eta_0  \left( 1 - \frac{ 1 - \exp(-\Delta t/\tau) }{\Delta t/\tau} \right).
\end{equation}
With the above formulation, irreversible effects can manifest only if sufficient time elapses for them actually to take place. Parameter $\tau$ represents the material characteristic time while $\eta_0$ acts as a scaling factor.

\subsection{Non-associative flow rule}
The class of materials target of this model do not exhibit stress-hardening deformation in tensile stress states, but rather loss of stiffness and consequent failure and initiation of fracture, for this reason inelastic flow in tensile stress state is completely neglected and the material response in such loading scenario is handled by the phase-field fracture formulation.

To avoid any interference between the damage model and the irreversible deformation one, the flow rule \eqref{eq:associative flow rule} is modified to allow plastic flow only in compressive stress states and those are identified using the pressure stress invariant, leveraging the explicit definition of the plastic multiplier defined in equation \eqref{explicit definition plastic multiplier}
\begin{equation}
    \begin{aligned}
    \dot{\gamma} = \begin{cases}
            \eta \langle F(\bm\sigma, \bm{A}) \rangle \quad & \text{if} \quad p>0 \\
            0 & \text{otherwise}
        \end{cases}
    \end{aligned}
\end{equation}

%% file: 04_Industrial_application.tex
\section{Application to industrial component} 

\begin{figure}[hbt!]
    \centering
    \includegraphics[width=0.5\textwidth]{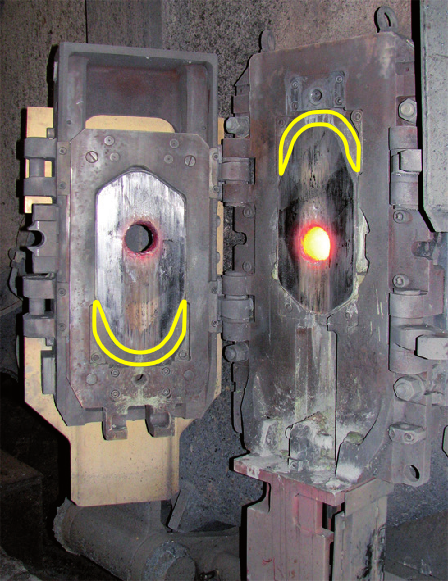}
    \caption{Field picture of a slide gate. The component highlighted in yellow is part of the plate fixture, and it is usually referred to as a horseshoe. Picture liberally taken from the website of \cite{VesuviusLadleGates}.}
    \label{fig:slide_gate_preparation}
\end{figure}

% What are slide gate plates ?
The model presented in this contribution has been applied to studying an industrial refractory product in operational conditions. The choice for the application case fell on the slide gate plate, an important component of the slide gate, the mechanism used to open and close molten metal flux from vessels commonly known as tundish. Figure \ref{fig:slide_gate_preparation} shows a field picture of an open slide gate that allows a clear view of the plates.

% Why Slide Gate Plates have the shape they have

Fractures of slide gate plates are a normal occurrence that happens by design in specific zones of the component due to the thermal gradients acting on the piece during contact with the molten metal. It can be said, without exaggeration, that no plate exists that has been put in operation in a slide gate and has not cracked.

The fractures happening because of normal operating conditions are due to a thermal gradient developing between the hot inner bore and the cooler outer areas of the plate. These normal fractures are meant not to cross the throttling path, the zone responsible for holding the column of liquid steel. Thus, the plate's shape is such that its inhomogeneous thermal expansion, constrained by the plate fixtures, leads only to safe fractures. 
\ifnum \censura=1 {Example pictures of typical fractures on slide gate plates can be found in \cite{Lee_Thomas_Kim_2016, Tang_Lu_Li_Zhu_Yi_Liu_Eckert_2023}.
} \else {Figure \ref{fig:cracked_plates} shows some examples of plates after normal operation.
} \fi

\ifnum \censura=1 {} \else {
    \begin{figure}[h!]
        \centering
        \includegraphics[width=\textwidth]{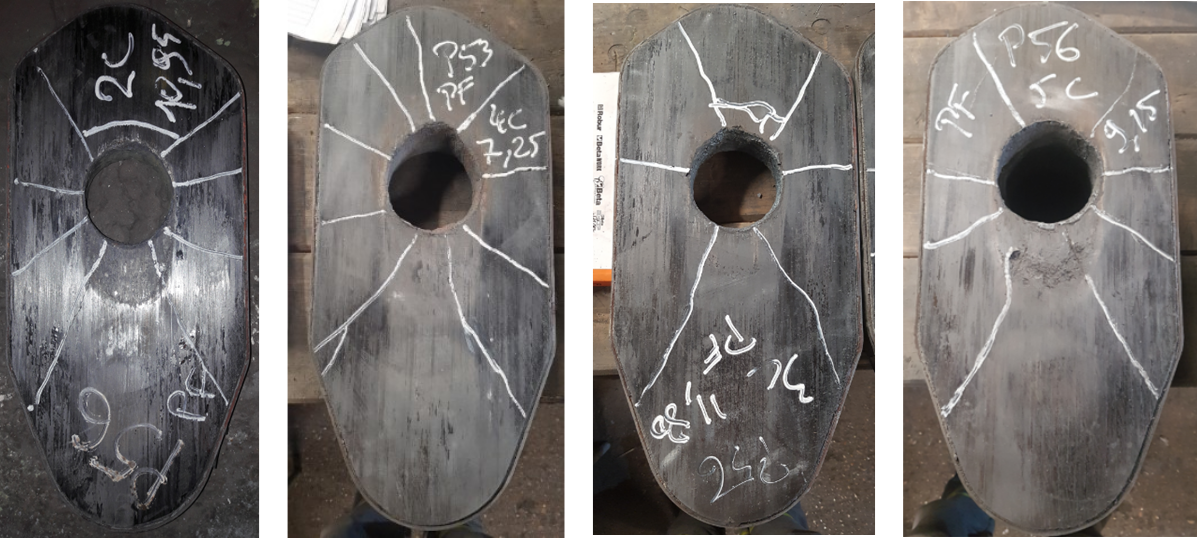}
        \caption{Field pictures of slide gate plates after usage, courtesy of Vesuvius. The fracture paths are highlighted in white.}
        \label{fig:cracked_plates}
    \end{figure}
} \fi

\subsection{Boundary Conditions}
The numerical model of the Slide Gate Plate presented here is based on the plane stress hypothesis. To save computational resources, the geometry is simplified with the assumption of one plane of symmetry, although real slide gate plates are not symmetric. This simplification is judged not to alter the phenomena of interest.

\begin{figure}[hbt!]
    \centering
    \includegraphics[width=0.8\textwidth]{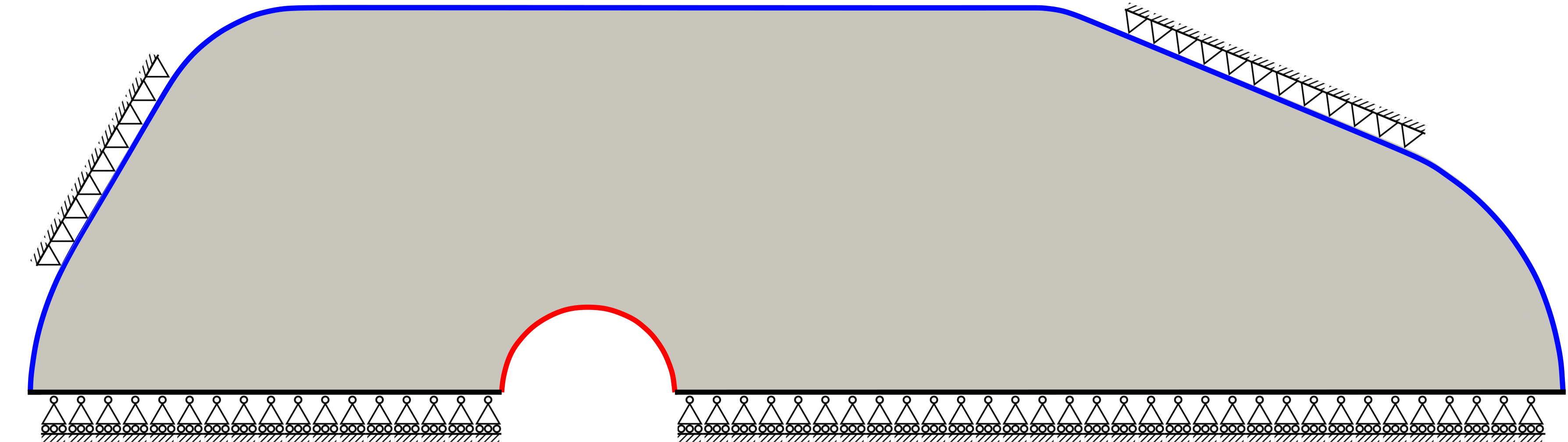}
    \caption{Schematic representation of the thermo-mechanical boundary conditions for the chosen case study. The plate bore is subjected to heating convective boundary conditions, while the external edge is subjected to cooling. The symmetry boundary is considered adiabatic and constrained in displacement orthogonal to it, while two portions of the external boundary are considered pinned.}
    \label{fig:SGP_CNS_boundary_conditions}
\end{figure}

Figure \ref{fig:SGP_CNS_boundary_conditions} represents the thermo-mechanical boundary conditions:
\begin{itemize}
    \item the plate bore is subjected to a convective heat flux condition of expression:
    \begin{equation}
        q_{\partial\Omega} = h_{steel} ~ (T_{steel} - T_{\partial\Omega}),
    \end{equation}
    where the convective coefficient $h_{steel}$ is set at a value of $50 \cdot 10^3 \unit[per-mode = symbol]{\watt\per\square\meter\per\kelvin}$; while the steel temperature is set at $1560 \unit{\degreeCelsius}$;
    \item the plate outer boundary is subjected to a convective heat flux condition of expression:
    \begin{equation}
        q_{\partial\Omega} = h_{outer} ~ (T_{outer} - T_{\partial\Omega}) ,
    \end{equation}
    where the convective coefficient $h_{outer}$ is set at a value of $50 \unit[per-mode = symbol]{\watt\per\square\meter\per\kelvin}$; while the outer temperature is set at $24 \unit{\degreeCelsius}$;
    \item the symmetry boundary condition is set as a constrain for the plate not to move in the direction orthogonal to the symmetry boundary face;
	\item the conditions of installation of the plate are simplified by imposing null displacement on the zones of the outer shell of the plate corresponding to the zones of contact of the refractory piece with its fixture.
\end{itemize}

The values for the convection coefficients have been chosen following the available information in the related scientific literature and industrial practice (see, for example, \cite{Lee_Thomas_Kim_2016} for deriving the coefficient starting from thermodynamic considerations).

For simplicity, the effect of the contact with the fixtures is not considered in the conduction model. In the actual slide gate, the plates are not rigidly constrained as one of the fixture components, usually referred to as horseshoe due to its shape, is spring-loaded and compresses the plate on its plane. Figure \ref{fig:slide_gate_preparation} shows a field picture of a slide gate during installation; the horseshoe is highlighted in yellow. The boundary conditions of this study are thus more severe than the ones in actual operation. This is not judged problematic as it implies that the model will give conservative results, the prediction of a fracture does not necessarily mean an actual fracture on the real piece while the opposite is less likely to happen.

Actual slide gate plates have a steel band along the perimeter.
The model does not consider the steel band's presence at the plate's edge. This choice is due to the uncertainties related to the mechanical contact between the plate and the steel band due to the manufacturing process of the plates themselves. The band's placement around the refractory is operated with mechanical interference by preheating the steel band and letting it cool down on the plate. The resulting mechanical contact is not perfect, and it is difficult to represent numerically. Also, in this case, not representing the steel band is judged to lead to more critical conditions for the plate as its thermal expansion as the band acts compressing the plate and mitigating the tensile stresses due to the inhomogeneous thermal expansion.

\subsection{Thermal convection-conduction model}
The thermal problem is solved as transient: the temperature evolution in the domain depends on the simulation time. The variational form of the transient thermal problem in weak form reads:
\begin{equation}
\label{eq:variational thermal pb}
\begin{aligned}
    &\int_\Omega ~ \rho ~ c_p ~ \dot{T} ~ \delta T ~ d\Omega +
     \int_\Omega K ~ \nabla T \cdot \nabla  \delta T ~ d\Omega + \\
  + &\int_{\partial\Omega} h_{steel} ~ (T_{steel} - T_{\partial\Omega}) ~ \delta T ~ d\partial\Omega +  \\
  + &\int_{\partial\Omega} h_{outer} ~ (T_{outer} - T_{\partial\Omega}) ~ \delta T ~ d\partial\Omega = 0,
\end{aligned}
\end{equation}
where $\delta T$ is the test function.

For the sake of clarity, it is important to underline that the thermal field's solution is also influenced by the initiation and propagation of fracture, as the thermal conduction properties of the continuum are affected by the insurgence of discontinuities. This effect is neglected here for simplicity, but it could be considered by introducing a suitable degradation function for the thermal conduction tensor in the thermal partial different equation. 

\subsection{Thermal expansion mechanical model}

The mechanical problem is expressed in variational weak form as:
\begin{equation}
\label{eq:variational mech pb}
    \int_\Omega g(d)~\bm\sigma \cdot \delta\bm\varepsilon ~ d\Omega = 0,
\end{equation}
where the deformation measure $\delta\bm\varepsilon$ is an algebraic manipulation of the displacement test function $\delta \bm{u}$ according to the kinematic hypothesis:
\begin{equation}
    \bm\varepsilon = \frac{1}{2} \left( \nabla \bm u + \nabla \bm u^T \right) .
\end{equation}
The stress field is computed from the elastic strain field using the expression:
\begin{gather}
    \bm\sigma (\bm{u}) = \lambda(T) \tr[\bm\varepsilon_e(\bm{u})] + 2 \mu(T) \bm\varepsilon_e(\bm{u}) , \\
    \bm\varepsilon_e(\bm{u}) = \bm\varepsilon(\bm{u}) - \bm\varepsilon_p - \alpha(T)(T-T_0) \bm{I},
\end{gather}
where $\lambda$ and $\mu$ are the Lam\`{e} parameters, $\alpha$ is the thermal expansion coefficient, $T_0$ is the reference temperature, here assumed at $24 \unit{\degreeCelsius}$, and $\bm{I}$ is the identity matrix. $\lambda$, $\mu$ and $\alpha$ are assumed as thermal-dependent properties.

\subsection{Phase field simulation of fracture on Slide Gate Plate}

The main advantage of applying the phase field approach to fracture to the case study at hand is the possibility to identify the location of crack initiation and the crack propagation path without any special treatment other than the solution of the damage partial different equation over the domain. The solution process is not straightforward for several reasons:
\begin{itemize}
    \item the strain energy split and the crack driving force formulation introduce a non-linearity in the coupled mechanical-damage problem; even leveraging the hybrid formulation of \citep{Ambati_Gerasimov_De_Lorenzis_2015}, which allows a linear formulation for the mechanical problem,
    the coupled solution for both fields has to be found iteratively for every load increment;
    \item since the fracture behaviour is modelled as quasi-brittle, consistent fracture propagation happens in a single load increment, and the magnitude of such increment affects the convergence of the solution;
    \item the mesh size in the zone interested by the fracture propagation plays a fundamental role, effectively influencing the structure's stiffness.
\end{itemize}
The issues described above have been progressively addressed when setting up the simulation. All the computations have been performed using the finite element library FEniCS with custom code developed by the author to define the variational formulations, the constitutive model and the solution procedure.

\subsection{Adaptive time stepping}

The problem at hand is transient, with physical time playing a role in the temperature evolution and the development of irreversible deformations. Additionally, since the phenomenon of interest is the initiation and propagation of fracture, it is advantageous to solve for small time steps and output the solution in the increments that involve damage evolution. To achieve such a result, the time step is controlled with an ad-hoc algorithm whose logic involves the control of a scalar quantity representing the cumulative increase of damage in a single increment. As a scalar measure of damage over the entire domain, the choice fell on the formula from \cite{Borden_Hughes_Landis_Anvari_Lee_2016}:
\begin{equation}
    \label{eq:Borden_Hughes_Landis_Anvari_Lee_2016}
    \psi_{d} = \int_\Omega \frac{G_c}{4~\ell} \left( d^2 + 4 \ell^2 \left| \nabla d \right|^2 \right) d\Omega.
\end{equation}
Given the enforcement of irreversibility of damage, $\psi_{d}$ is monotonically increasing.

The time step control algorithm evaluates $\psi_{d}$ for the computed solution and compares the value with that of the last accepted solution. If the increase in $\psi_{d}$ is less than a certain threshold (here taken as $60\%$), the solution at the increment is accepted, the time step is raised, and the simulation is continued. Otherwise, the solution is rejected, the time step lowered, and the incremental solution recalculated. The time step is additionally kept within a maximum and minimum interval. The minimum time step is here taken as \num{e-6}\unit{\second}.

\subsection{Mesh refinement algorithm}
The phase field representation of fracture is founded on the idea of representing discontinuities as concentrated gradients of a scalar field. The zones of concentration of gradients necessitate sufficiently small meshes for a correct representation with the finite element method. The disadvantage of this requirement lies in the high computational cost associated with using fine meshes on the whole computational domain. The pre-allocation of zones with refined mesh has the disadvantage that it implies previous knowledge about the location of fracture initiation and propagation. This contribution proposes an ad-hoc adaptive mesh refinement strategy to overcome this limitation.

The mesh refinement algorithm presented here leverages the functionalities offered by the library FEniCS for the operations related to the splitting of the cells and the definition of the necessary matrices representing the equations of the physical problem over the tasselled domain. The user only has to specify the conditions for flagging the mesh elements. It is then possible to split the flagged elements and define the problem over the new mesh, bringing the information from the coarse one.

The mesh refinement happens dynamically during the solution procedure in the implementation presented here. At every increment, the solution of the multi-physics problem is operated in a staggered approach by first solving the thermal problem, then solving the mechanical one, and then flagging the mesh using the mesh refinement conditions. In the case mesh refinement is judged necessary due to a certain number of elements being flagged, mesh refinement is operated. After checking the mesh refinement condition and the eventual mesh refinement, which also entails the projection of the quantities already calculated on the coarse mesh on the new one, the damage problem is solved. The procedure is repeated iteratively until convergence is reached on an energetic measure. This means that the mesh refinement procedure can be applied multiple times in the solution of a certain load increment. This ensures that the representation of fracture happens on a coherent mesh size, independently from the initial mesh.

The mesh refinement conditions above are designed to ensure the best phase-field fracture representation. This objective is pursued by combining different strategies available in the related literature \citep{Freddi_Mingazzi_2022, Hirshikesh_Schneider_Nestler_2023}. The mesh elements are flagged if:
\begin{itemize}
    \item the crack driving force $\mathcal{H}$ is over a certain threshold; this condition ensures that mesh refinement is operated in the zones where the fracture has not yet happened, but it is likely to happen due to the high value of the crack driving force;
    \item the damage is over a certain threshold; since the damage varies continuously, this condition ensures a gradual decrease in mesh size around the zones of already developed fractures;
    \item the element is greater than a certain minimum size, this ensures that the refinement procedure is eventually stopped and the refined mesh in the interested zones has the desired size linked to the length scale parameter.
\end{itemize}

\subsection{Geometrical and material parameters}

This section reports the geometrical and material parameters used for the simulations shown in this chapter. It is reminded that the values reported here are plausible values of the same order of magnitude as real experimental values and part designs but do not correspond to any real material or product.

The slide gate plate geometry analyzed in this chapter has an overall size of $440$ \unit{\milli\meter} by $110$ \unit{\milli\meter}, the two inclined surfaces have angles of $60^{\circ}$ and $20^{\circ}$, the major fillet radius are of $75$ \unit{\milli\meter} while the minor ones are of $45$ \unit{\milli\meter}. The bore is positioned at $160$  mm from the ``short end'' of the plate (the one identified by the inclined surface of $60^{\circ}$) and has a radius of $25$ \unit{\milli\meter}. The plate thickness is irrelevant as the simulation is conducted in plane stress conditions without evaluating reaction forces or other quantities that need to be integrated over the third dimension.

Table \ref{tab:thermal_dependent_material_params} reports the thermal dependent material parameters used in the simulation, similar values can be found for reference in the literature \citep{Blond_Schmitt_Hild_Blumenfeld_Poirier_2005, Warchal_Andre_DeBastiani_Guillo_Huger_Martelli_Mazerat_Romero-Baivier_2017, Tang_Lu_Li_Zhu_Yi_Liu_Eckert_2023, Lee_Thomas_Kim_2016}.

\begin{table}[h]
\centering
\begin{tabular}{cccccc}
\toprule
Temperature   & Emod    & $\alpha$   & $K_{cond}$  & $c_p$        & $p_c$   \\
\unit{\degreeCelsius} & \unit{\giga\pascal} & \num{e-6}\unit{\per\kelvin} & \unit{\watt\per\meter\per\kelvin} & \unit{\joule\per\kg\per\kelvin} & \unit{\mega\pascal} \\
\midrule
20            & 50      &            & 9           & 1000         & 350     \\
200           &         & 5          &             &              &         \\
400           & 40      & 6          &             &              &         \\
600           & 30      & 7          & 8           & 1300         & 360     \\
800           &         &            &             &              & 330     \\
900           &         &            & 7           & 1400         &         \\
1000          &         &            &             &              & 300     \\
1200          & 20      &            &             &              & 200     \\
\bottomrule
\end{tabular}
\caption{Thermal dependent material parameters used in the simulations reported in this chapter.}
\label{tab:thermal_dependent_material_params}
\end{table}

The model is fully defined with the additional thermal-invariant material parameters. The elastic behaviour of the material is fully defined by specifying the Poisson ratio of $0.1$. The thermal behaviour is completed with the density specification of $3000$ \unit{\kilogram\per\meter^3}. The plasticity model requires the definition of the remaining six parameters for the BP yield surface ($p_c$ is assumed to have a thermal dependency) and of the hardening parameters: $M=1.6$, $m=2$, $\beta=0.75$, $\gamma=0.7$, $\alpha=0.1$, $c=5$ \unit{\mega\pascal}, $\chi=10$ \unit{\giga\pascal}, $\delta=500$. The visco-plastic parameter $\eta_0$ is taken as $5 \cdot 10^6$ while the material characteristic time is assumed to be 0.05 \unit{\second}. 

The fracture length scale parameter for the phase field model has been set to $0.5$ \unit{\milli\meter} coherently to represent fractures visible with the naked eye, while the specific fracture energy is assumed to be a thermal-dependent property. In the lack of specific data for this property, a value in the range of $150 \pm 20$ \unit{\joule\per\meter^2} was set at reference temperature and a value in the range $100 \pm 20$ \unit{\joule\per\meter^2} was set at $600$ \unit{\degreeCelsius} with linear interpolation inside the interval and constant approximation outside it.

\subsection{Model predictions}
\begin{figure}[hbt!]
    \centering
    \includegraphics[width=\textwidth]{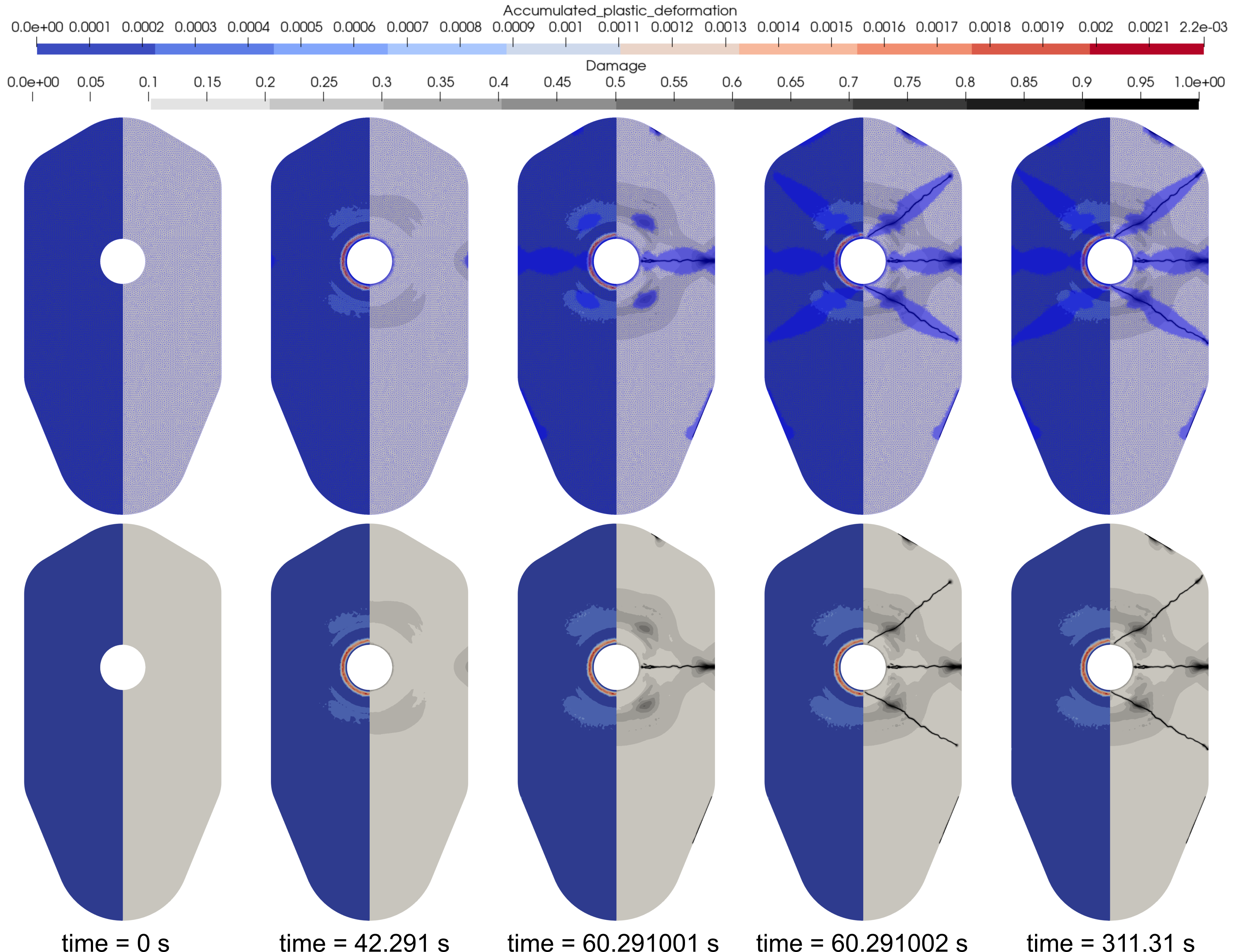}
    \caption{Damage and accumulated plastic deformation in slide gate plate geometry subjected to thermal-shock conditions.}
    \label{fig:240528T1725_SGP_TEPD_summary}
\end{figure}

\begin{figure}[hbt!]
    \centering
    \includegraphics[width=\textwidth]{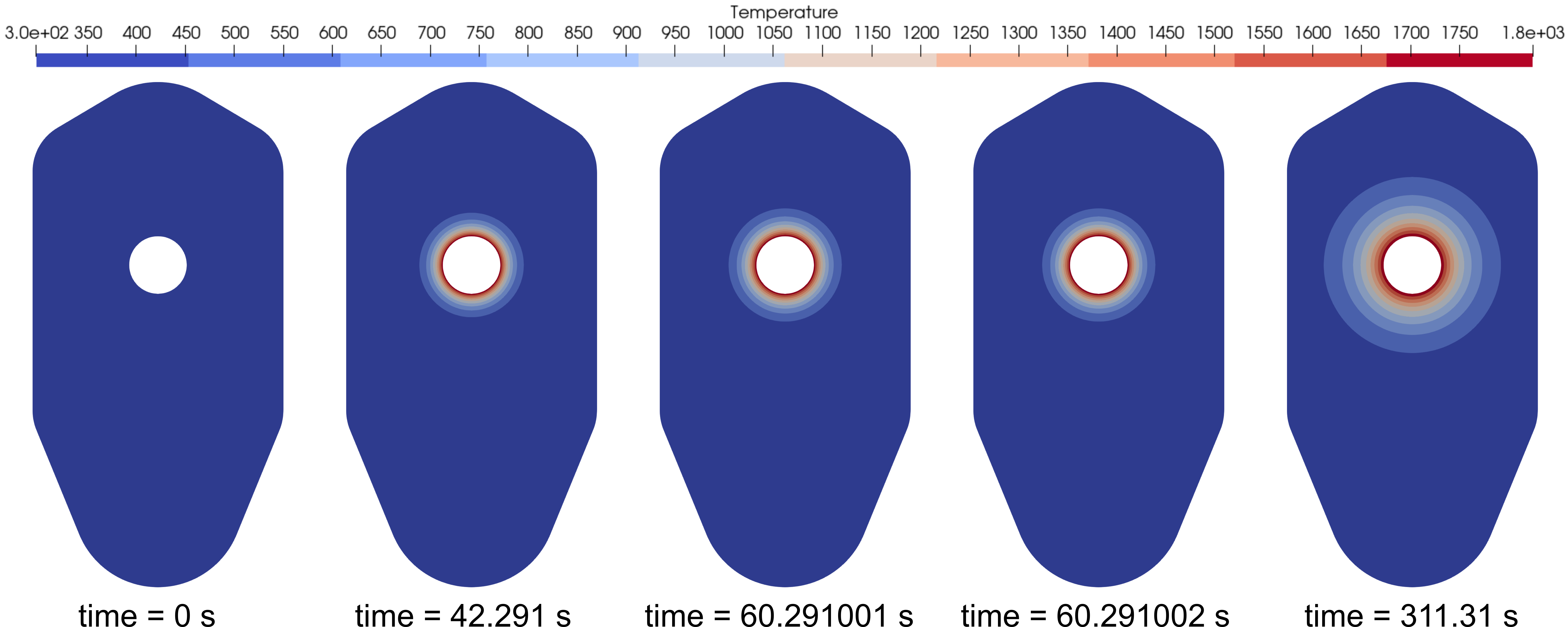}
    \caption{Temperature evolution in slide gate plate geometry subjected to thermal-shock conditions.}
    \label{fig:240528T1725_SGP_TEPD_temp}
\end{figure}

The results are presented in a compact form in figure \ref{fig:240528T1725_SGP_TEPD_summary}. The left-hand side of the plate reports the accumulated plastic deformation field, while the right-hand side reports the damage field. The top row of results presents the mesh, while the bottom one reports the same results without the mesh representation. Figure \ref{fig:240528T1725_SGP_TEPD_temp} presents the corresponding temperature field.

The solution at simulation time $42.291$ \unit{\second} clearly shows the initial development of irreversible deformations in the zone around the bore. This is consistent with observing the high gradient in the temperature field. Even though the plate bore is at $1800$ \unit{\degreeCelsius}, the temperature is unchanged in half of the plate's smaller section. The thermal expansion of the zone of the plate around the bore is thus constrained by the rest of the plate's material, which results in yielding under compression around the bore.

The solution at time $42.291$ \unit{\second} already shows damage concentration in a zone at the outer edge of the plate, closest to the bore. At time $60.291001$ \unit{\second}, the first fracture propagates from the aforementioned zone of damage concentration towards the bore. The tensile stresses that cause such fracture are induced for mechanical equilibrium by compression action from the plate's outer portion towards the zone around the bore. At this simulation time, it is already possible to observe two zones of concentration of damage at a certain distance from the bore, almost symmetric with respect to the already propagated fracture. Interestingly, these zones correspond to the zones of development of irreversible deformations.

At time $60.291002$ \unit{\second}, other two fractures propagate from the aforementioned zones in two opposing directions: towards the bore and the plate's outer edge. These fractures are supposed to be due to the thermal expansion of the part of the plate that is not directly affected by the action of the constraints. As a portion of the plate does not have a direct constraint to thermal expansion, its displacement causes tensile-shear stresses that lead to fracture. 

After the development of the three characteristic fractures, further heating of the plate does not lead to relevant events in the considered simulation time of circa $300$ \unit{\second}. This is consistent with the observation that damage evolution in these thermo-mechanical conditions is due to the transient thermal gradient. The plate is subjected to less critical stresses as the temperature homogenises due to conduction. The instants of fracture propagation also confirm the transient nature of the interested phenomena. It is interesting to observe how propagation is not captured even if the time step is brought to the minimum allowed value of 1 micro-second, and fracture still happens in just one step.

\ifnum \censura=1 {} \else {
Figure \ref{fig:IMG_1324} shows three details for the characteristic fractures in a real slide gate plate. Naked eye observation reveals the zone of major crack opening on the plate surface. Interestingly, both the ``inclined'' fractures (top and bottom rows of the figure) do not seem to propagate up to the bore and the outer edge, at least not to the naked eye, supporting the observations about the location of fracture initiation asserted during the discussion of the simulation results.

\begin{figure}[hbt!]
    \centering
    \includegraphics[width=.8\textwidth]{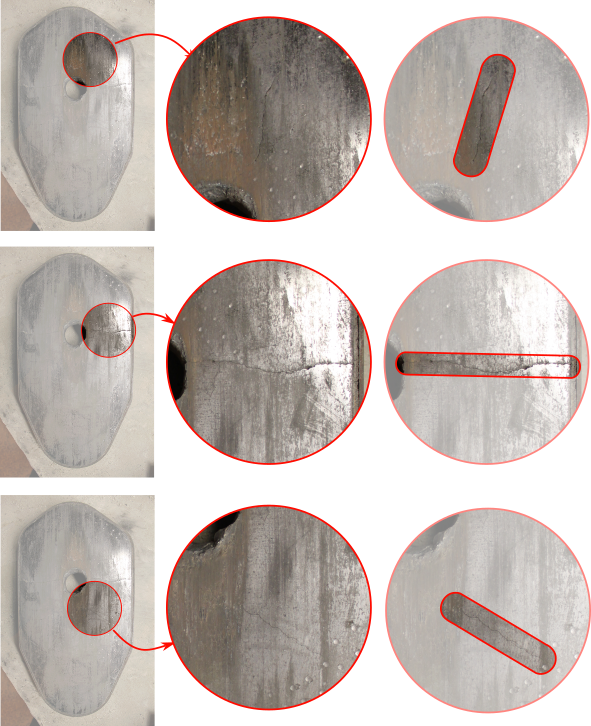}
    \caption{Details of the three characteristic fractures on a slide gate plate. Picture gently provided by Vesuvius.}
    \label{fig:IMG_1324}
\end{figure}
} \fi

%% file: 05_Conclusion.tex
\section{Conclusion}

This contribution presents a novel material model for ceramic refractory materials designed to represent the asymmetric tension-compression behaviour of these materials to quantitatively predict failures happening in real industrial parts, in the whole range of working temperatures.

The design of the material model starts from the analysis of the desired features identified based on the experimental observations. Specifically two behaviours are of interest and are the targets for the modelling: inelastic flow and fracture propagation. This approach leads to the identification of two main models that can represent the interested behaviours: the continuum mechanics theory of plasticity and the phase-field fracture damage model. 

Using the two aforementioned modelling paradigms, the model is complete with the choice of the Bigoni-Piccolroaz yield function, isotropic thermo-mechanical hardening laws for the yield function parameters and the formulation for the phase field crack driving force.

The model is thus implemented in the open-source Finite Element framework FEniCS and used to simulate the transient phenomenon of fracture initiation and propagation in Slide Gate Plates in operational conditions. In order to save computational resources, without compromising the solution quality, the implementation also features ad-hoc adaptive algorithms for the control of the time step and the refinement of the mesh.

The predictions of the model show good qualitative accordance with real parts despite the simplifying assumptions. The three fractures analyzed in the presented study case are the characteristic ones that are always expected to appear in real pieces, so the simplifications that have been made are judged satisfactory to lead to realistic representation and the model is judged to be a useful tool for the design of new industrial refractory components.

%% file: 06_Appendix.tex
\section{Model recap}

\begin{tcolorbox}[colback=white, colframe=black, coltitle=white, title= The thermal problem: conduction and convection]
    \begin{equation*}
    \begin{aligned}
        &\int_\Omega ~ \rho ~ c_p ~ \dot{T} ~ \delta T ~ d\Omega +
         \int_\Omega K ~ \nabla T \cdot \nabla  \delta T ~ d\Omega + \\
      + &\int_{\partial\Omega} h_{steel} ~ (T_{steel} - T_{\partial\Omega}) ~ \delta T ~ d\partial\Omega +  \\
      + &\int_{\partial\Omega} h_{outer} ~ (T_{outer} - T_{\partial\Omega}) ~ \delta T ~ d\partial\Omega = 0
    \end{aligned}
    \end{equation*}
\end{tcolorbox}

\begin{tcolorbox}[colback=white, colframe=black, coltitle=white, title= The mechanical problem: deformation and stress]
    \begin{equation*}
        \int_\Omega (1-d)^2~\mathbb{E}_0(T)~ \cdot \biggr[ \bm\varepsilon(\bm{u}) - \bm\varepsilon_p - \alpha (T-T_0) ~ \bm{I} \biggr] \cdot ~ \bm\varepsilon(\delta\bm{u}) ~ d\Omega = 0
    \end{equation*}
\end{tcolorbox}

\begin{tcolorbox}[colback=white, colframe=black, coltitle=white, title= The plasticity problem: irreversible deformations]
    \begin{tcolorbox}[colback=white, colframe=gray, coltitle=white, title= Yield function definition]
        \begin{gather*}
            F(\bm\sigma) = F(p,q,\theta) = f(p) + \frac{q}{g(\theta)} \\
            \Phi = \frac{p+c}{p_c+c} \\[1ex]
            f(p) = \begin{cases*}
                - M p_c \sqrt{(\Phi-\Phi^m)\left[2(1-\alpha)\Phi + \alpha \right]}\\
                +\infty \quad \Phi\notin[0;1]
            \end{cases*} \\[1ex]
            g(\theta)^{-1} = cos \left[ \beta \frac{\pi}{6} - \frac{1}{3} arccos(\gamma cos(3\theta)) \right]
        \end{gather*}
    \end{tcolorbox}

    \begin{tcolorbox}[colback=white, colframe=gray, coltitle=white, title= Internal variable]
        \begin{equation*}
            \bar\varepsilon_p = \int_{0}^{t} || \dot{\bm\varepsilon}_p || dt
        \end{equation*}
    \end{tcolorbox}
    
    \begin{tcolorbox}[colback=white, colframe=gray, coltitle=white, title= Thermo-mechanical hardening]
        \begin{align*}
            p_c &= p_{cT} + p_{cM} \\
            p_{cM} &= \dfrac{\chi}{1+\delta ~ \bar\varepsilon_p} ~ \bar\varepsilon_p \\[2ex]
            p_{cT} &= \text{table}(T)
        \end{align*}
    \end{tcolorbox}

    \begin{tcolorbox}[colback=white, colframe=gray, coltitle=white, title= Flow rule]
        \begin{equation*}
            \dot{\bm\varepsilon}_p = \eta \langle F(\bm\sigma) \rangle ~ 
            \dfrac{\partial F}{\partial \bm\sigma} \quad \text{if} \quad p>0
        \end{equation*}
    \end{tcolorbox}
    
\end{tcolorbox}

\begin{tcolorbox}[colback=white, colframe=black, coltitle=white, title= The damage problem]
    \begin{tcolorbox}[colback=white, colframe=gray, coltitle=white, title= Fracture evolution]
        \begin{equation*}
            \int_\Omega \biggr[ -2(1-d) ~ \delta d ~ \dfrac{\mathcal{H}}{G_c} + \dfrac{1}{\ell} d \delta d + \ell ~ \nabla d \cdot \nabla \delta d \biggr] d\Omega=0
        \end{equation*}
    \end{tcolorbox}
    
    \begin{tcolorbox}[colback=white, colframe=gray, coltitle=white, title= Fracture driving force]
        \begin{gather*}
            \mathcal{H} = \max_{s\in[0,t]} \left\lbrace \psi_{+}(\bm\varepsilon)/G_c \right\rbrace \\[1ex]
            \psi(\bm\varepsilon) = \psi_{-}(\bm\varepsilon) + g(d) ~ \psi_{+}(\bm\varepsilon) \\[1ex]
            \psi_{\pm} = \frac{\lambda}{2} \langle tr[\bm\varepsilon] \rangle_{\pm}^2 + \mu tr[\bm\varepsilon_\pm^2] \\[1ex]
            \bm\varepsilon = \sum_i \langle \bm\varepsilon \rangle_\pm e_i \otimes e_i
        \end{gather*}
    \end{tcolorbox}
\end{tcolorbox}

\section{Implementation algorithm}
The present model is implemented in the finite element framework FEniCS which allows a simple representation and resolution of PDEs in computer code. 
The code targets thermo-mechanical problems and a staggered strategy is used to solve the multi-physics problem as suggested by \cite{Miehe_Welschinger_Hofacker_2010}:

\begin{enumerate}
    \item the thermal problem is solved first and the thermo-elastic effect is neglected;
    \item the displacement problem is solved using the fields of temperature, plastic strain and damage computed at the previous iteration;
    \item the plastic strain is updated with the stress tensor value resulting from the updated displacement and the plastic strain of the previous time step;
    \item the damage problem is solved using the updated displacement field;
    \item the solution process is iterated until convergence on an energy measure is reached.
\end{enumerate}

Compared to the monolithic scheme, the staggered scheme described above is particularly advantageous for its speed, simplicity of implementation and numerical robustness, especially for the cases in which the damage progresses rapidly. The limitation of this method is related to the time step size, which has to be small enough to assure convergence and accuracy.

Additional computational cost is avoided by implying a hybrid formulation \citep{Ambati_Gerasimov_De_Lorenzis_2015} for the strain energy split: the split is used only to define the crack driving force, while the whole elastic strain energy is degraded by the damage field in the displacement problem. The crack closure effect can still be captured using the additional condition proposed by \citep{Ambati_Gerasimov_De_Lorenzis_2015}:
\begin{equation}
    \forall \bm{x} : \Psi_D < \Psi_R \implies d = 0.
\end{equation}
which basically consists of setting the damage to a null value in the material points subjected to compression.

The visco-plastic model presented in this contribution involves the local solution of a non-linear system of equations for the update of the state at a certain material point. Differently from standard plasticity models, the relationship ruling the stress computation does not involve implicit expressions, in particular the constrain relative to the null value of the yield function does not have to be enforced. This relaxation allows the use of a simple and effective iterative approach for the computation of the stress state. This kind of approach was first proposed by \cite{Zienkiewicz_Cormeau_1974} for standard plasticity models, with the argument that the imposition of the standard Karush–Kuhn–Tucker conditions is more a mathematical formalism that a true physically-based condition. 

Following the nomenclature used in standard return mapping algorithms, the elastic predictor stress state is first computed using the imposed strain increment, the resulting stress is then inserted into the explicit expression for the plastic multiplier, which is in turn used to compute the plastic strain at the current increment. In the case in which the plastic multiplier is not null, the newly computed plastic strain affects the stress value so that the computation has to be iterated until convergence. In order to improve accuracy, an inner loop is operated at every time step and the plastic strain is calculated at every increment as the previously calculated plastic strain plus an increment which depends both on the previous and the converged time steps:
\begin{gather}
    \bm\varepsilon_p^{n+1} = \bm\varepsilon_p^{n} + \Delta \bm\varepsilon^{p n} \\[1ex]
    \Delta \bm\varepsilon^{p n} = \frac{1}{2}(\dot{\bm\varepsilon}_p^n + \dot{\bm\varepsilon}_p^{n+1}) \Delta t
\end{gather}

Leveraging this unconstrained regularization of the standard constrained plasticity problem, the usual return mapping algorithm used to calculate stress is substituted by the explicit procedure described in the box \textbf{Iterative algorithm for regularized plasticity}.

\begin{tcolorbox}[colback=white, colframe=gray, coltitle=white, title=Iterative algorithm for regularized plasticity]
    \begin{algorithm}[H]
        \KwInput{
        \begin{itemize}
            \item Previous step stress $\bm\sigma^n$, strain $\bm\varepsilon^n$ and load increment $\Delta\bm\varepsilon^{n+1}$ imposed by the global solution algorithm
            \item State of the material point at time step $n$ described by the local internal variables: $\bm\varepsilon_p^n$, $\bar{\varepsilon}_p^n$, $\gamma^n$, $p_c^n$
        \end{itemize}
        }
        \KwOutput{
        \begin{itemize}
            \item Updated stress $\bm\sigma^{n+1}$ to be returned to the global equilibrium algorithm
            \item Updated values for the internal variables: $\bm\varepsilon_p^{n+1}$, $\bar{\varepsilon}_p^{n+1}$, $\gamma^{n+1}$, $p_c^{n+1}$
        \end{itemize}
        }
        % \KwData{Testing set $x$ \\ \quad \\}
        % \DontPrintSemicolon

        Initialize $\bm\sigma^{n+1} = \bm\sigma^{n}$, $d^{n+1} = d^{n}$, $\bm\varepsilon^{n+1} = \bm\varepsilon^{n} + \Delta\bm\varepsilon^{n+1}$ and $\bm\varepsilon_p^{n+1} = \bm\varepsilon_p^{n}$ \\[2ex]
        Calculate $\bm\sigma^{n+1} = g(d^{n+1})\, \mathbb{D}\left[ \bm\varepsilon^{n+1} - \bm\varepsilon_p^{n+1} \right]$ \\[2ex]
        Calculate $\dot{\bm\varepsilon}_p^{n+1} = \eta \langle F(\bm\sigma^{n+1}) \rangle \dfrac{\partial F^{n+1}}{\partial \bm\sigma^{n+1}} $ \\[2ex]
        Calculate $\Delta \bm\varepsilon^{n+1}_p = \frac{1}{2}(\dot{\bm\varepsilon}_p^n + \dot{\bm\varepsilon}_p^{n+1}) \Delta t$ \\[2ex]
        Update $\bm\varepsilon_p^{n+1} = \bm\varepsilon_p^{n} + \Delta \bm\varepsilon^{p n}$ \\[2ex]
        Calculate $\Delta\bar{\varepsilon}_p^{n+1} = \eta \Delta t \dfrac{1}{2} \left[ \langle F(\bm\sigma^n) \rangle \Big{|}\Big{|}\dfrac{\partial F^n}{\partial \bm\sigma^n}\Big{|}\Big{|} + \langle F(\bm\sigma^{n+1}) \rangle \Big{|}\Big{|}\dfrac{\partial F^{n+1}}{\partial \bm\sigma^{n+1}}\Big{|}\Big{|} \right] $ \\[2ex]
        Update $ \bar{\varepsilon}_p^{n+1} = \bar{\varepsilon}_p^n + \Delta\bar{\varepsilon}_p^{n+1} $ \\[2ex]
        Update $p_c^{n+1}$ with $\bar{\varepsilon}_p^{n+1}$ \\[2ex]
        Solve the displacement problem for $\bm{u}^{n+1}(\bm\varepsilon_p^{n+1} , d^{n+1})$ \\[2ex]
        Solve the damage problem for $d^{n+1}$ \\[2ex]
        Iterate steps from 2 to 10 until convergence is reached on elastic energy
    \end{algorithm}
\end{tcolorbox}

\cite{Zienkiewicz_Cormeau_1974}, who first proposed this algorithm, raised the point that the iteration process is found to be not needed in practical applications since the magnitude of the time step is already limited by stability conditions. In their study, the authors propose analytical solutions for stable time steps for some plastic constitutive models. The stability limits for the time steps involve both the elasticity and plasticity properties of the constitutive model and especially the regularization viscosity term $\eta$. In the present study, a comprehensive study of the stability as a function of the time step is not relevant as the time step is already limited by the iterative procedure for the solution of the phase-field fracture problem.